\documentstyle[10pt,epsf,epsfig,dp_delphititle,oldlfont,amssymb]{dp_delphi}
\makeindex
\def\DpPaperGroup{EP-PH}
\def\DpPaperRef{2005-028}
\def\DpDate{5 July 2005}
\def\DpAuthors{DELPHI Collaboration}
\def\DpTitle{{
Search for {\bf $\eta_b$} in two-photon collisions at LEP II with the 
DELPHI detector }}
\def\DpSubmit{(Accepted by Phys. Lett. B)}
\def\DpComment{ }
\def\DpEMail{  }
\def\jpsi{\hbox{$J{\kern-0.24em}/{\kern-0.14em}\psi$}}
\def\ev#1#2{\hbox{#1e{\kern-0.10em}V{\kern-0.30em}/{\kern-0.14em}$#2$}}

\def\bqt#1#2\eqt{\begin{equation}\label{#1}%
{#2}\end{equation}\noindent}
\def\bln#1#2\eln{\begin{equation}\label{#1}%
\eqalign{#2}\end{equation}\noindent}
\def\brlist{\begin{thebibliography}{99}}

\def\erlist{\end{thebibliography}}

\begin{document}
\makeatletter
\newcount\@tempcntc
\def\@citex[#1]#2{\if@filesw\immediate\write\@auxout{\string\citation{#2}}\fi
  \@tempcnta\z@\@tempcntb\m@ne\def\@citea{}\@cite{\@for\@citeb:=#2\do
    {\@ifundefined
       {b@\@citeb}{\@citeo\@tempcntb\m@ne\@citea\def\@citea{,}{\bf ?}\@warning
       {Citation `\@citeb' on page \thepage \space undefined}}%
    {\setbox\z@\hbox{\global\@tempcntc0\csname b@\@citeb\endcsname\relax}%
     \ifnum\@tempcntc=\z@ \@citeo\@tempcntb\m@ne
       \@citea\def\@citea{,}\hbox{\csname b@\@citeb\endcsname}%
     \else
      \advance\@tempcntb\@ne
      \ifnum\@tempcntb=\@tempcntc
      \else\advance\@tempcntb\m@ne\@citeo
      \@tempcnta\@tempcntc\@tempcntb\@tempcntc\fi\fi}}\@citeo}{#1}}
\def\@citeo{\ifnum\@tempcnta>\@tempcntb\else\@citea\def\@citea{,}%
  \ifnum\@tempcnta=\@tempcntb\the\@tempcnta\else
   {\advance\@tempcnta\@ne\ifnum\@tempcnta=\@tempcntb \else \def\@citea{--}\fi
    \advance\@tempcnta\m@ne\the\@tempcnta\@citea\the\@tempcntb}\fi\fi}
 
\makeatother
\begin{titlepage}
\pagenumbering{roman}
\CERNpreprint{\DpPaperGroup}{\DpPaperRef} 
\date{{\small\DpDate}} 
\title{\DpTitle} 
\address{\DpAuthors} 
\begin{shortabs} 
\noindent
The pseudoscalar meson $\eta_b$ has been searched for in two-photon interactions 
at LEP~II. The data sample corresponds to a total integrated luminosity 
of \mbox{617 pb$^{-1}$} at centre-of-mass energies ranging 
from 161 to 209~GeV.
Upper limits at a confidence level of 95\% on the product
$\Gamma_{\gamma\gamma}(\eta_b)\times$BR$(\eta_b)$ are 
190, 470 and 660~ eV\textit{/$c^2$} for the  
$\eta_b$ decaying into 4, 6 and 8 charged particles, respectively.

\end{shortabs}
\vfill
\begin{center}
\DpSubmit \ \\ 
\DpComment \ \\
\DpEMail \ \\
\end{center}
\vfill
\clearpage
\headsep 10.0pt
\addtolength{\textheight}{10mm}
\addtolength{\footskip}{-5mm}
\begingroup
%
\newcommand{\DpName}[2]{\hbox{#1$^{\ref{#2}}$},\hfill}
\newcommand{\DpNameTwo}[3]{\hbox{#1$^{\ref{#2},\ref{#3}}$},\hfill}
\newcommand{\DpNameThree}[4]{\hbox{#1$^{\ref{#2},\ref{#3},\ref{#4}}$},\hfill}
\newskip\Bigfill \Bigfill = 0pt plus 1000fill
\newcommand{\DpNameLast}[2]{\hbox{#1$^{\ref{#2}}$}\hspace{\Bigfill}}
%
\footnotesize
\noindent
\DpName{J.Abdallah}{LPNHE}
\DpName{P.Abreu}{LIP}
\DpName{W.Adam}{VIENNA}
\DpName{P.Adzic}{DEMOKRITOS}
\DpName{T.Albrecht}{KARLSRUHE}
\DpName{T.Alderweireld}{AIM}
\DpName{R.Alemany-Fernandez}{CERN}
\DpName{T.Allmendinger}{KARLSRUHE}
\DpName{P.P.Allport}{LIVERPOOL}
\DpName{U.Amaldi}{MILANO2}
% ...
\normalsize
\endgroup
\titlefoot{Department of Physics and Astronomy, Iowa State
     University, Ames IA 50011-3160, USA
    \label{AMES}}
\titlefoot{Physics Department, Universiteit Antwerpen,
     Universiteitsplein 1, B-2610 Antwerpen, Belgium \\
     \indent~~and IIHE, ULB-VUB,
     Pleinlaan 2, B-1050 Brussels, Belgium \\
     \indent~~and Facult\'e des Sciences,
     Univ. de l'Etat Mons, Av. Maistriau 19, B-7000 Mons, Belgium
    \label{AIM}}
\titlefoot{Physics Laboratory, University of Athens, Solonos Str.
     104, GR-10680 Athens, Greece
    \label{ATHENS}}
\titlefoot{Department of Physics, University of Bergen,
     All\'egaten 55, NO-5007 Bergen, Norway
    \label{BERGEN}}
\titlefoot{Dipartimento di Fisica, Universit\`a di Bologna and INFN,
     Via Irnerio 46, IT-40126 Bologna, Italy
    \label{BOLOGNA}}
\titlefoot{Centro Brasileiro de Pesquisas F\'{\i}sicas, rua Xavier Sigaud 150,
     BR-22290 Rio de Janeiro, Brazil \\
     \indent~~and Depto. de F\'{\i}sica, Pont. Univ. Cat\'olica,
     C.P. 38071 BR-22453 Rio de Janeiro, Brazil \\
     \indent~~and Inst. de F\'{\i}sica, Univ. Estadual do Rio de Janeiro,
     rua S\~{a}o Francisco Xavier 524, Rio de Janeiro, Brazil
    \label{BRASIL}}
\titlefoot{Coll\`ege de France, Lab. de Physique Corpusculaire, IN2P3-CNRS,
     FR-75231 Paris Cedex 05, France
    \label{CDF}}
\titlefoot{CERN, CH-1211 Geneva 23, Switzerland
    \label{CERN}}
\titlefoot{Institut de Recherches Subatomiques, IN2P3 - CNRS/ULP - BP20,
     FR-67037 Strasbourg Cedex, France
    \label{CRN}}
\titlefoot{Now at DESY-Zeuthen, Platanenallee 6, D-15735 Zeuthen, Germany
    \label{DESY}}
\titlefoot{Institute of Nuclear Physics, N.C.S.R. Demokritos,
     P.O. Box 60228, GR-15310 Athens, Greece
    \label{DEMOKRITOS}}
\titlefoot{FZU, Inst. of Phys. of the C.A.S. High Energy Physics Division,
     Na Slovance 2, CZ-180 40, Praha 8, Czech Republic
    \label{FZU}}
\titlefoot{Dipartimento di Fisica, Universit\`a di Genova and INFN,
     Via Dodecaneso 33, IT-16146 Genova, Italy
    \label{GENOVA}}
\titlefoot{Institut des Sciences Nucl\'eaires, IN2P3-CNRS, Universit\'e
     de Grenoble 1, FR-38026 Grenoble Cedex, France
    \label{GRENOBLE}}
\titlefoot{Helsinki Institute of Physics and Department of Physical Sciences,
     P.O. Box 64, FIN-00014 University of Helsinki, 
     \indent~~Finland
    \label{HELSINKI}}
\titlefoot{Joint Institute for Nuclear Research, Dubna, Head Post
     Office, P.O. Box 79, RU-101 000 Moscow, Russian Federation
    \label{JINR}}
\titlefoot{Institut f\"ur Experimentelle Kernphysik,
     Universit\"at Karlsruhe, Postfach 6980, DE-76128 Karlsruhe,
     Germany
    \label{KARLSRUHE}}
\titlefoot{Institute of Nuclear Physics PAN,Ul. Radzikowskiego 152,
     PL-31142 Krakow, Poland
    \label{KRAKOW1}}
\titlefoot{Faculty of Physics and Nuclear Techniques, University of Mining
     and Metallurgy, PL-30055 Krakow, Poland
    \label{KRAKOW2}}
\titlefoot{Universit\'e de Paris-Sud, Lab. de l'Acc\'el\'erateur
     Lin\'eaire, IN2P3-CNRS, B\^{a}t. 200, FR-91405 Orsay Cedex, France
    \label{LAL}}
\titlefoot{School of Physics and Chemistry, University of Lancaster,
     Lancaster LA1 4YB, UK
    \label{LANCASTER}}
\titlefoot{LIP, IST, FCUL - Av. Elias Garcia, 14-$1^{o}$,
     PT-1000 Lisboa Codex, Portugal
    \label{LIP}}
\titlefoot{Department of Physics, University of Liverpool, P.O.
     Box 147, Liverpool L69 3BX, UK
    \label{LIVERPOOL}}
\titlefoot{Dept. of Physics and Astronomy, Kelvin Building,
     University of Glasgow, Glasgow G12 8QQ
    \label{GLASGOW}}
\titlefoot{LPNHE, IN2P3-CNRS, Univ.~Paris VI et VII, Tour 33 (RdC),
     4 place Jussieu, FR-75252 Paris Cedex 05, France
    \label{LPNHE}}
\titlefoot{Department of Physics, University of Lund,
     S\"olvegatan 14, SE-223 63 Lund, Sweden
    \label{LUND}}
\titlefoot{Universit\'e Claude Bernard de Lyon, IPNL, IN2P3-CNRS,
     FR-69622 Villeurbanne Cedex, France
    \label{LYON}}
\titlefoot{Dipartimento di Fisica, Universit\`a di Milano and INFN-MILANO,
     Via Celoria 16, IT-20133 Milan, Italy
    \label{MILANO}}
\titlefoot{Dipartimento di Fisica, Univ. di Milano-Bicocca and
     INFN-MILANO, Piazza della Scienza 2, IT-20126 Milan, Italy
    \label{MILANO2}}
\titlefoot{IPNP of MFF, Charles Univ., Areal MFF,
     V Holesovickach 2, CZ-180 00, Praha 8, Czech Republic
    \label{NC}}
\titlefoot{NIKHEF, Postbus 41882, NL-1009 DB
     Amsterdam, The Netherlands
    \label{NIKHEF}}
\titlefoot{National Technical University, Physics Department,
     Zografou Campus, GR-15773 Athens, Greece
    \label{NTU-ATHENS}}
\titlefoot{Physics Department, University of Oslo, Blindern,
     NO-0316 Oslo, Norway
    \label{OSLO}}
\titlefoot{Dpto. Fisica, Univ. Oviedo, Avda. Calvo Sotelo
     s/n, ES-33007 Oviedo, Spain
    \label{OVIEDO}}
\titlefoot{Department of Physics, University of Oxford,
     Keble Road, Oxford OX1 3RH, UK
    \label{OXFORD}}
\titlefoot{Dipartimento di Fisica, Universit\`a di Padova and
     INFN, Via Marzolo 8, IT-35131 Padua, Italy
    \label{PADOVA}}
\titlefoot{Rutherford Appleton Laboratory, Chilton, Didcot
     OX11 OQX, UK
    \label{RAL}}
\titlefoot{Dipartimento di Fisica, Universit\`a di Roma II and
     INFN, Tor Vergata, IT-00173 Rome, Italy
    \label{ROMA2}}
\titlefoot{Dipartimento di Fisica, Universit\`a di Roma III and
     INFN, Via della Vasca Navale 84, IT-00146 Rome, Italy
    \label{ROMA3}}
\titlefoot{DAPNIA/Service de Physique des Particules,
     CEA-Saclay, FR-91191 Gif-sur-Yvette Cedex, France
    \label{SACLAY}}
\titlefoot{Instituto de Fisica de Cantabria (CSIC-UC), Avda.
     los Castros s/n, ES-39006 Santander, Spain
    \label{SANTANDER}}
\titlefoot{Inst. for High Energy Physics, Serpukov
     P.O. Box 35, Protvino, (Moscow Region), Russian Federation
    \label{SERPUKHOV}}
\titlefoot{J. Stefan Institute, Jamova 39, SI-1000 Ljubljana, Slovenia
     and Laboratory for Astroparticle Physics,\\
     \indent~~Nova Gorica Polytechnic, Kostanjeviska 16a, SI-5000 Nova Gorica, Slovenia, \\
     \indent~~and Department of Physics, University of Ljubljana,
     SI-1000 Ljubljana, Slovenia
    \label{SLOVENIJA}}
\titlefoot{Fysikum, Stockholm University,
     Box 6730, SE-113 85 Stockholm, Sweden
    \label{STOCKHOLM}}
\titlefoot{Dipartimento di Fisica Sperimentale, Universit\`a di
     Torino and INFN, Via P. Giuria 1, IT-10125 Turin, Italy
    \label{TORINO}}
\titlefoot{INFN,Sezione di Torino and Dipartimento di Fisica Teorica,
     Universit\`a di Torino, Via Giuria 1,
     IT-10125 Turin, Italy
    \label{TORINOTH}}
\titlefoot{Dipartimento di Fisica, Universit\`a di Trieste and
     INFN, Via A. Valerio 2, IT-34127 Trieste, Italy \\
     \indent~~and Istituto di Fisica, Universit\`a di Udine,
     IT-33100 Udine, Italy
    \label{TU}}
\titlefoot{Univ. Federal do Rio de Janeiro, C.P. 68528
     Cidade Univ., Ilha do Fund\~ao
     BR-21945-970 Rio de Janeiro, Brazil
    \label{UFRJ}}
\titlefoot{Department of Radiation Sciences, University of
     Uppsala, P.O. Box 535, SE-751 21 Uppsala, Sweden
    \label{UPPSALA}}
\titlefoot{IFIC, Valencia-CSIC, and D.F.A.M.N., U. de Valencia,
     Avda. Dr. Moliner 50, ES-46100 Burjassot (Valencia), Spain
    \label{VALENCIA}}
\titlefoot{Institut f\"ur Hochenergiephysik, \"Osterr. Akad.
     d. Wissensch., Nikolsdorfergasse 18, AT-1050 Vienna, Austria
    \label{VIENNA}}
\titlefoot{Inst. Nuclear Studies and University of Warsaw, Ul.
     Hoza 69, PL-00681 Warsaw, Poland
    \label{WARSZAWA}}
\titlefoot{Fachbereich Physik, University of Wuppertal, Postfach
     100 127, DE-42097 Wuppertal, Germany \\
\noindent
{$^\dagger$~deceased}
    \label{WUPPERTAL}}
\addtolength{\textheight}{-10mm}
\addtolength{\footskip}{5mm}
\clearpage
\headsep 30.0pt
\end{titlepage}
%
\pagenumbering{arabic} 
\setcounter{footnote}{0} %
\large
\section{Introduction}
Two-photon collisions are very useful in searching for the formation
of  pseudoscalar mesons
with $J^{PC}=0^{-+}$. The high energy and high luminosity of LEP II
are additional motivations to look for the $b \bar b$ pseudoscalar
quarkonium state $\eta_b$ which has not yet been discovered~\cite{ALEPH,L3}.

Its mass, $m_{\eta_b}$,
 is estimated by several theoretical models~\cite{MASS}.
It should lie below  that of the $\Upsilon$ vector meson 
($m_{\Upsilon}$=9.46~GeV\textit{/$c^2$}) and
the mass shift, $\Delta m = m_{\Upsilon}- m_{\eta_b}$, is estimated to 
be in the range 10 to 130~MeV\textit{/$c^2$}.

The cross-section for two-photon resonance $R$ formation with C=+1 
$${\mathrm e^+}{\mathrm e^-}\rightarrow {\mathrm e^+}{\mathrm e^-}
\gamma^*\gamma^*
\rightarrow {\mathrm e^+}{\mathrm e^-} R$$
 is given by~\cite{SIGMA}
$$\sigma({\mathrm e^+}{\mathrm e^-}\rightarrow {\mathrm e^+}{\mathrm e^-}R)=
\int \sigma_{\gamma\gamma\rightarrow \eta_b} dL_{\gamma\gamma}(W^2), $$
with the cross-section
  
$$
 \sigma_{\gamma\gamma\rightarrow \eta_b}(W^2,q_1^2,q_2^2) = 8\pi\,(2J_R+1)
  \cdot \Gamma_{\gamma\gamma}(R)
 \cdot  F^2(q_1^2,q_2^2)\cdot{\Gamma_R\over
  (W^2-m_R^2)^2+m_R^2\Gamma_R^2}.
$$
Here $L_{\gamma\gamma}(W^2)$ is the two-photon luminosity function, $W$ is 
the two-photon centre-of-mass energy, $q_1^2$ and $q_2^2$ 
are the squares of the virtual-photon four-momenta.
The resonance $R$ is characterised by its spin $J_R$, mass $m_R$, total 
width $\Gamma_R$ and its two-photon partial width 
$\Gamma_{\gamma\gamma}(R)$. In ``quasi-real'' ($q^2\sim 0$) photon
interactions, the form factor $F^2(q_1^2,q_2^2)$ is constant and 
can be taken to be unity.
 
To compute the $\eta_b$ production cross-section, the
partial width $\Gamma_{\gamma\gamma}(\eta_b)$ must be known.  
Theoretical estimates~\cite{GGG} predict it to be in 
the range 260 to 580~eV\textit{/$c^2$}. 
Setting $m_{\eta_b}$ to 9.4~GeV\textit{/$c^2$} ~leads to an expected
 production cross-section 
$\sigma({\mathrm e^+}{\mathrm e^-}\rightarrow {\mathrm e^+}{\mathrm e^-}
\eta_b)$ of \mbox{0.14 to 0.32 pb} at $\sqrt s$=200 GeV.

Most of the observations of $\eta_c$ decays have been to four charged 
particles, both pions and kaons ~\cite{PDG}. Hence the $\eta_b$ has been 
similarly searched
for in 4, 6 and 8 charged particle final states. The expected backgrounds come
from the $\gamma\gamma \rightarrow q \bar q$ processes and 
the $\gamma\gamma \rightarrow \tau^+\tau^-$ channel.

From the ALEPH experiment, upper limits on 
$\Gamma_{\gamma\gamma}(\eta_b)\times$BR$(\eta_b)$ ~\cite{ALEPH} are :
\begin{center}
$\Gamma_{\gamma\gamma}(\eta_b)\times$BR$(\eta_b\rightarrow$ 4 charged particles)
 $<$ 48~eV\textit{/$c^2$}, \\
$\Gamma_{\gamma\gamma}(\eta_b)\times$BR$(\eta_b\rightarrow$ 6 charged particles)
 $<$ 132~eV\textit{/$c^2$}.
\end{center}
The L3 experiment, looking for 
$\eta_b$ in the decay modes $\eta_b\rightarrow$ 
$K^+$$K^-$$\pi^0$, ~$\pi^+$$\pi^-$$\eta$, 
\mbox{2, 4 and 6} charged particles (only or associated with one $\pi^0$), 
observes 6 candidate events with 2.5 background events expected. 
This corresponds to a combined  upper limit on 
$\Gamma_{\gamma\gamma}(\eta_b)\times$BR$(\eta_b)$~\cite{L3}:
\begin{center}
$\Gamma_{\gamma\gamma}(\eta_b)\times$BR$(\eta_b\rightarrow$ analysed channels)
 $<$ 200~eV\textit{/$c^2$}. 
\end{center}

\bigskip
In this paper we report on the search for $\eta_b$ 
in the reaction
$${\mathrm e^+}{\mathrm e^-}\rightarrow {\mathrm e^+}{\mathrm e^-}
\gamma^*\gamma^*
\rightarrow {\mathrm e^+}{\mathrm e^-} \eta_b$$
with $\eta_b$ decaying into the following final states: 
\begin{center}
$\eta_b \rightarrow 4\pi^{\pm}({\mathrm K}^{\pm})$, \\
$\eta_b \rightarrow 6\pi^{\pm}({\mathrm K}^{\pm})$, \\
$\eta_b \rightarrow 8\pi^{\pm}({\mathrm K}^{\pm}).$
\end{center}
Here the charged K's in parentheses indicate that a pair of pions may be 
replaced by a pair of kaons with net zero strangeness.

\section{Experimental procedure}
The analysis presented here is based on the data taken with the DELPHI 
detector~\cite{DELPHI1,DELPHI2} in 1996-2000, covering 
a range of centre-of-mass energies from 161 to 209 GeV
(average centre-of-mass energy: 195.7 GeV).
The selected data set corresponds to the period when the Time Projection 
Chamber (TPC) was fully 
operational thus ensuring good particle reconstruction.
This requirement reduces the integral luminosity for the analysis to
\mbox{617 ${\mathrm p}{\mathrm b}^{-1}$}.
  
For quasi-real photon interactions, the scattered ${\mathrm e}^{\pm}$ 
are emitted at very small polar angles. Hence there is no requirement on 
detecting them. 

The ${\mathrm e}^+{\mathrm e}^- \rightarrow {\mathrm e}^+{\mathrm e}^-\eta_b$
candidate events are selected by requiring final states with 4, 6 or 8 tracks
with zero net charge. 
Charged-particle tracks in the detector are accepted 
if the following criteria are met:
\begin{itemize}
\item particle transverse momentum $p_T \ >$ 150~MeV\textit{/$c$};
\item impact parameter of a track transverse to the beam axis
$\Delta_{xy} \ <$ 0.5 cm;
\item impact parameter of a track along the beam axis
$\Delta_{z} \ <$ 2 cm;
\item polar angle of a track 
$10^{\circ} \ < \ \theta \ < \ 170^{\circ}$;
\item track length $l \ >$ 30 cm;
\item relative error of the track momentum $\Delta p/p \ <$ 30\%.
\end{itemize}

No ${\mathrm K}^0_S$ reconstruction is attempted on each track pair.
The identification of other neutral particles is made using calorimeter 
information.
The calorimeter clusters which are not associated 
to charged-particle tracks are combined to form the signals from 
the neutral particles ($\gamma, \ \pi^0, \ {\mathrm K}^0_L,$ n).
A minimum measured
energy of 1 GeV for showers in the electromagnetic
calorimeters and 2 GeV  in the hadron calorimeters
is required.
   
The selection of candidate events is achieved by applying 
the following criteria: 
\begin{itemize}
\item no particle is identified as an electron or a muon by the
 standard lepton-identification algorithms~\cite{DELPHI3};
\item no particle is identified as a proton by the
 standard identification algorithm~\cite{DELPHI3};
\item there are no electromagnetic showers
with energy $E_{shower}>1 $ GeV or converted $\gamma$'s
with energy $E_{\gamma}>0.2 $ GeV in the event.  
\end{itemize}
To ensure that no particle from the $\eta_b$ decay
has escaped detection, the square of the total transverse momentum 
of charged particles , $(\sum \vec p_T)^2$, is required to be small.
The actual cut value is estimated  from a Monte Carlo sample 
of $\eta_b$ events produced  in $\gamma\gamma$ interactions.
In this simulation the kinematical variables are generated using 
the algorithms developed 
by Krasemann et al.~\cite{VERMASEREN}. It is also assumed that
the production amplitude factorizes into the quasi-real transverse
photon flux and a covariant amplitude describing both the $\eta_b$
production and decay~\cite{DECAY}.
The $\eta_b \rightarrow$ (4, 6, 8) charged-particle decay processes are
assumed to be described by the phase-space momenta distribution.
The generated events are passed through  the standard 
DELPHI detector simulation and reconstruction programs~\cite{DELPHI2}.
The same selection criteria are applied on the simulated events as 
on the data.
Finally, an event is accepted on the basis of the trigger efficiency. 
Parametrized for a single track, as a function of its transverse momentum 
$p_T$, it ranges from 20\% for \mbox{$p_T <$ 0.5~GeV\textit{/$c$}}
to about 95\% at $p_T >$ 2~GeV\textit{/$c$}~\cite{TRIGGER}. 
Due to the high mass of the
$\eta_b$ \mbox{resonant} state and relatively large number of tracks in the final 
state, the overall trigger efficiency per event is 
about 93.6\%, 94.5\% and 94.6\% for events with 4, 6 and 8 charged particles, 
respectively.

Fig.1 shows, in the visible invariant-mass 
interval \mbox{8~GeV/$c^2$~$<W_{vis}<$~10~GeV/$c^2$}, 
the fraction of remaining events as a function of a cut, $P_T^2$, 
on $(\sum \vec p_T)^2$, for the 4 charged-particle channel. It decreases 
rapidly for $P_T^2$  $<$ 0.1~GeV$^2$\textit{/c$^2$}.
Hence to preserve the statistics, 4, 6 charged-particle events with  
$(\sum \vec p_T)^2$ up to 0.08~GeV$^2$\textit{/c$^2$} and  8 charged-particle 
events with $(\sum \vec p_T)^2$ up to 0.06~GeV$^2$\textit{/$c^2$} were kept.

\begin{figure}[htb]
\epsfxsize=\textwidth
\begin{center}
\mbox{\epsfig{figure=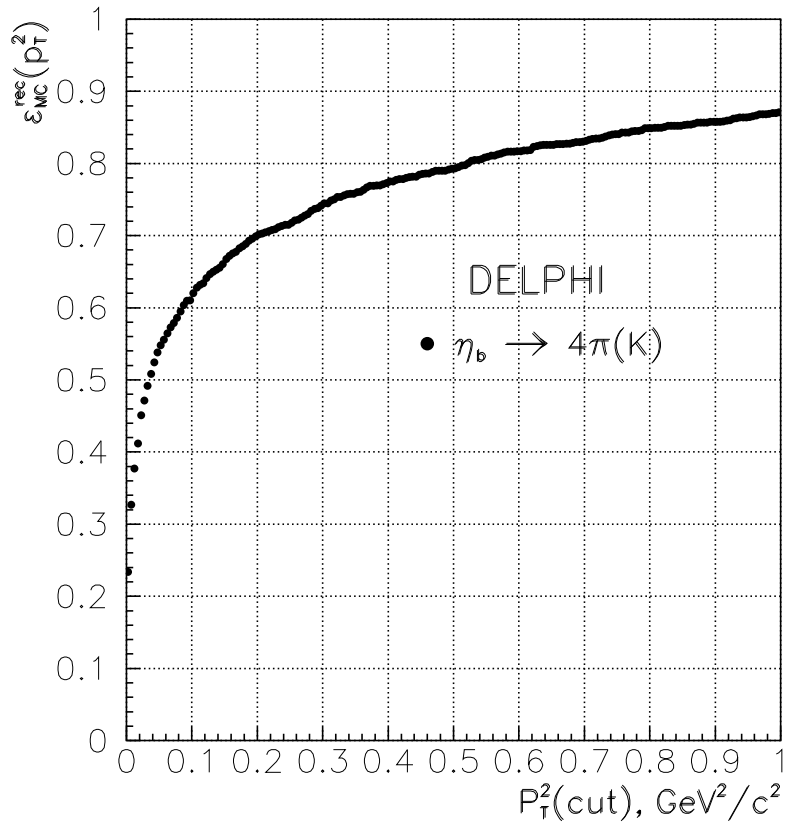,width=12. cm}}
\end{center}
\caption{Efficiency of selected $\eta_b$ Monte Carlo events
of the 4 charged-particle channel, as a function of
the cut $(\sum \vec p_T)^2$ $<$ $P_T^2$, in the $\eta_b$ search region:
 8~GeV\textit{/$c^2$}~$<W_{vis}<$~10~GeV\textit{/$c^2$}.}
\end{figure}
    
The $\pi$/K identification is based on the TPC dE/dx and 
RICH~\cite{RIBMEAN} measurements which are used both separately
and combined, in order to check the consistency, in a neural network-based 
algorithm~\cite{MACRIB}. 
In the $\eta_b$ search region defined as 
\mbox{8~GeV\textit{/$c^2$}~$<W_{vis}<$~10~GeV\textit{/$c^2$}}, the average
$K^{\pm}$ identification efficiency is about 54\% and the purity is 82\%.
The misidentification of charged pions as kaons is about 1.5\%. 
After application of the selection criteria and requiring 
$W_{vis}>$~5~GeV\textit{/$c^2$}, the 4, 6 and 8 charged-particle data samples 
contain 173, 328 and 113 events respectively.

The main background comes from inclusive  
$\gamma\gamma \rightarrow q \bar q$ channels. This background is estimated
using a Monte Carlo sample generated with the
PYTHIA 6.143 program~\cite{PYTHIA}.

The possible contamination of the  ${\mathrm e}^+{\mathrm e}^-\rightarrow 
{\mathrm e}^+{\mathrm e}^-\tau^+\tau^-$ process is given special 
attention. 
To reduce it in the  $\gamma  \gamma \rightarrow 4\pi$ channel
where it is most important, events of topology 1-3 with respect to the 
hemispheres
defined by the thrust axis computed in the $4\pi$ centre-of-mass system and 
with an invariant mass, in each hemisphere, smaller
than 1.8~GeV\textit{/$c^2$}, are discarded. 
Only (1.0$\pm$0.3)\% of $\eta_b$ events are eliminated by this cut.


The mass resolution in the search region has been estimated from the Monte 
Carlo sample of $\gamma\gamma \rightarrow q \bar q$ interactions.
It is about 200~MeV\textit{/$c^2$} FWHM for all topologies, as shown on Fig.2 
for the 4 charged-particle events. We have chosen to search for a possible 
signal in $\pm$ one mass resolution interval around the predicted mass 
of 9.4 GeV\textit{/$c^2$}.
\begin{figure}[htb]
\epsfxsize=\textwidth
\begin{center}
\mbox{\epsfig{figure=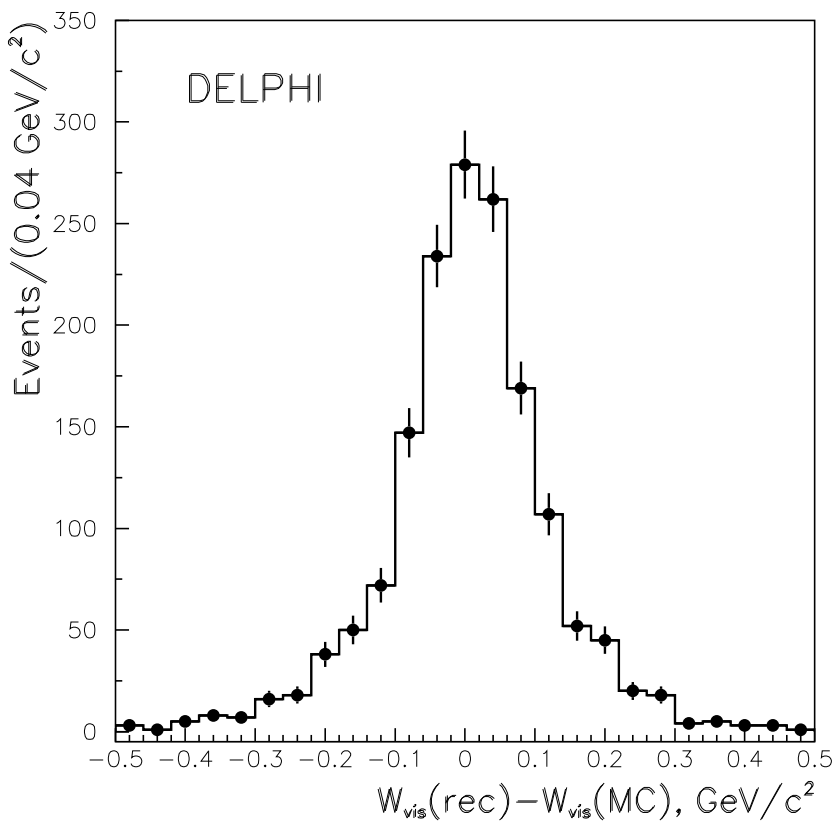,width=12. cm}}
\end{center}
\caption{Difference between reconstructed and generated 
$W_{vis}$ values for the selected 4 charged-particle events from the Monte 
Carlo $\gamma\gamma \rightarrow q \bar q$ sample,
in the $\eta_b$ search region.}
\end{figure}


\section{Results}
The visible invariant-mass spectra 
of the selected events are presented in Fig. 3.
When an event has an odd number of $K^{\pm}$, the kaon mass is assigned
sequentially to the other particles of opposite charge and the
$W_{vis}$ mass is simply taken as the average of the various mass combinations.
The resulting mass shift, averaged over the 4, 6 and 8 particle samples, is about 
120~MeV\textit{/$c^2$} in the $\eta_b$ search region.

The distributions are well reproduced by the 
$\gamma\gamma \rightarrow q \bar q$ Monte Carlo simulation. 
The $\eta_b$ candidates are expected to show up in the 9.2 
to 9.6~GeV\textit{/$c^2$} ~mass region.

\vskip10mm
\begin{figure*}[htb]
\epsfxsize=\textwidth
\begin{center}
\mbox{\epsfig{figure=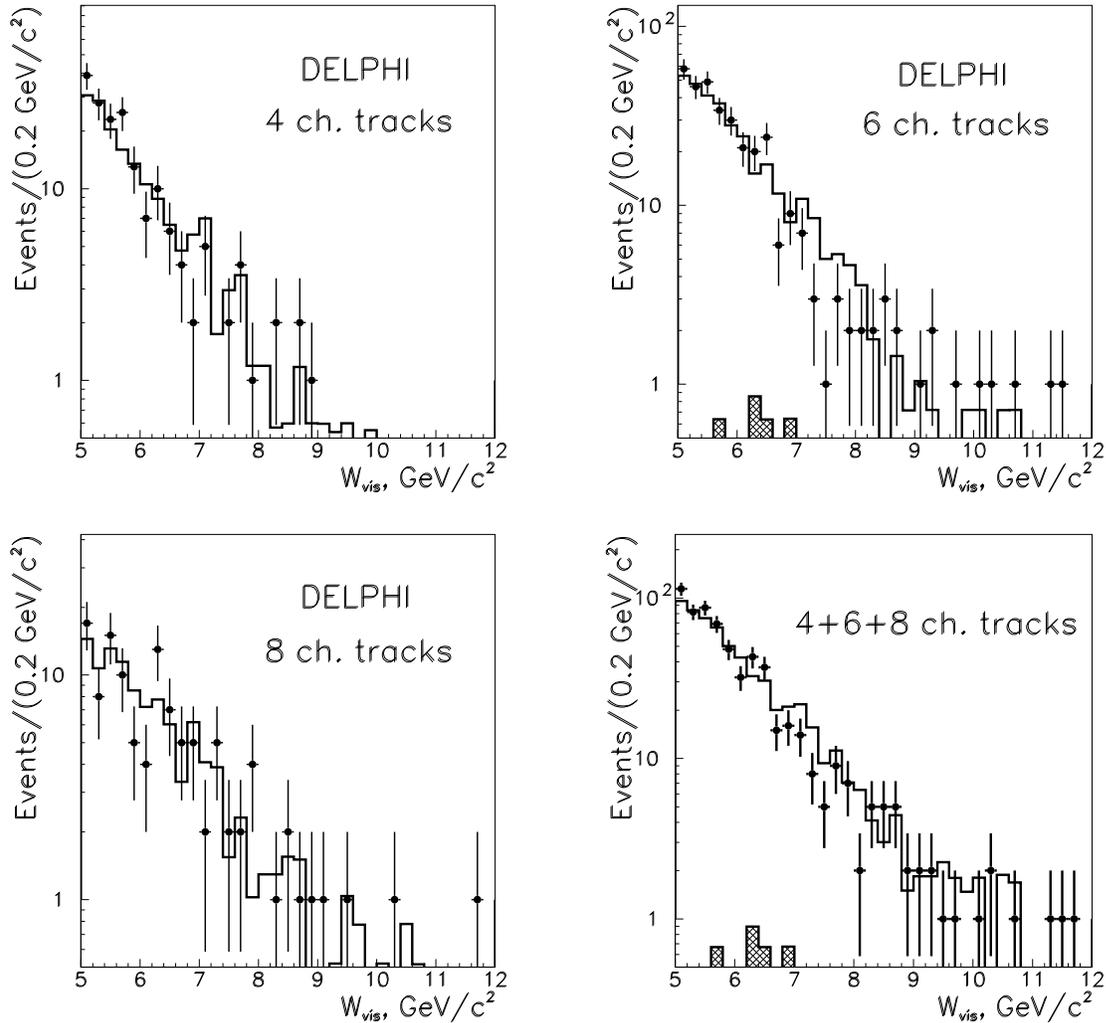,width=16. cm}}
\end{center}
\vspace*{-4.cm}
\caption{Invariant-mass distributions of selected events for 
4, 6 and 8 charged-particle final states. Points with error bars
are from the data; histograms present the expected number of background events
from the $\gamma\gamma \rightarrow q \bar q$ simulation;
shaded histograms correspond to the expected 
${\mathrm e}^+{\mathrm e}^-\rightarrow 
{\mathrm e}^+{\mathrm e}^-\tau^+\tau^-$ background.}
\end{figure*}

Table 1 gives the number of 4, 6 and 8 charged-particle events in the 
9.2 to 9.6 GeV\textit{/$c^2$} ~mass region,
together with the number of expected background events computed 
taking into account the overall reconstruction and selection efficiency. 
Among the 3 observed $\eta_b$ candidates only the event 
with 8 charged particles contains one identified kaon.
\begin{table}[htb]
\begin{center}
\begin{tabular}{|c|c|c|c|}
 \hline
   &
\multicolumn{3}{|c|}{$\eta_b$ decay modes}    \\
 &4 ch.tracks ($N_{bkg}$) &
6 ch.tracks ($N_{bkg}$) & 8 ch.tracks ($N_{bkg}$)\\
\hline 
 $N_{obs}$ (9.2~$<W_{vis}<$9.6~GeV\textit{/$c^2$}) & 0 $\quad$ (1.2) & 2  $\quad$ (1.1) &
 1  $\quad$  (1.5) \\
\hline
\hline
$N_{ev}$ & 3.9 & 5.7 & 4.1 \\
 (95\% C.L. upper limit) & & & \\
\hline
 overall efficiency & 5.9\% & 3.5\% & 1.8\% \\
\hline
$\Gamma_{\gamma\gamma}(\eta_b)\times$BR$(\eta_b)$, eV\textit{/$c^2$} & 190 & 470 & 660 \\
 (95\% C.L. upper limit) & & & \\
\hline
\end{tabular}
\caption{Number of observed 4, 6 and 8 charged-particle $\eta_b$ 
candidates ($N_{obs}$),
expected background events ($N_{bkg}$), 95\% C.L. upper limits for signal
events ($N_{ev}$), overall efficiency and 95\% C.L. upper limits on  
$\Gamma_{\gamma\gamma}(\eta_b)\times$BR$(\eta_b).$}
\end{center}
\end{table}
 
In the search for rare processes with a few observed events that may be 
compatible with background, an upper limit for the signal $S$ can be 
derived considering a Poisson process with a background $b$ and taking into
account uncertainties in the background and efficiencies~\cite{Zech}
$${\mathrm C}{\mathrm L}= 1-\frac{\int g(b) f(\varepsilon)\sum_{k=0}^{n}
P[k|(S\varepsilon+b)] d\varepsilon db}{\int g(b) \sum_{k=0}^{n} P(k|b) db}.$$
Here $P(k|x)$ is the Poisson probability of $k$ events being observed, when
$x$ are expected;
CL is a confidence level, $n$ is the number of observed events.
The probability-density functions for the background $g(b)$ and the efficiency
$f(\varepsilon)$ are assumed to be Gaussian and restricted to the range 
where $b$ and $\varepsilon$ are positive.

Upper limits at the 95\% confidence level were calculated for each channel and
a limit on
$\Gamma_{\gamma\gamma}(\eta_b)\times$BR$(\eta_b)$ could then be derived. 
The values are quoted in Table 1. 

We considered as main sources of systematic uncertainties:
the statistical error of the background, the generator used for the 
$\eta_b$ signal
and the theoretical uncertainties of the $\eta_b$ parameters. 
The limited statistics of our Monte Carlo event sample introduces relative 
uncertainties of 3\%, 5\%, 4\% for the channels with
4, 6 and 8 charged particles respectively.
To appreciate the influence of the generators, we have used PHOT02 
~\cite{ALEPH,PHOT02} which generates $\eta_b$ events decaying into two gluon-jets. 
The relative differences in efficiency are of 24\%, 11.4\% and 6.1\% for 
the 4, 6 and 8 charged particles channels.
Varying the $\eta_b$ mass within the range of 9.33 -- 9.45~GeV\textit{/$c^2$} 
~generates 
a relative uncertainty of 2.5\% on $N_{ev}$, for each considered $\eta_b$ 
decay channel. The three kinds of uncertainties were added quadratically to 
obtain the upper limits quoted in Table 1.

\section{Conclusions}
The pseudoscalar meson $\eta_b$ has been searched for through its decays to 
4, 6 and 8 charged-particles in two-photon interactions 
at LEP II.  The data sample corresponds to a total integrated luminosity 
of 617 pb$^{-1}$ collected at centre-of-mass energies ranging 
from 161 to 209~GeV.

Upper limits at a confidence level of 95\% on the product
$\Gamma_{\gamma\gamma}(\eta_b)\times$BR$(\eta_b)$ are 
190, 470 and 660~eV\textit{/$c^2$} for the  
$\eta_b \rightarrow$ (4, 6, 8) charged particle decays, respectively.

\subsection*{Acknowledgements}
\vskip 3 mm
 We are greatly indebted to our technical 
collaborators, to the members of the CERN-SL Division for the excellent 
performance of the LEP collider, and to the funding agencies for their

support in building and operating the DELPHI detector.\\
We acknowledge in particular the support of \\
Austrian Federal Ministry of Education, Science and Culture,
GZ 616.364/2-III/2a/98, \\
FNRS--FWO, Flanders Institute to encourage scientific and technological 
research in the industry (IWT), Belgium,  \\
FINEP, CNPq, CAPES, FUJB and FAPERJ, Brazil, \\
Czech Ministry of Industry and Trade, GA CR 202/99/1362,\\
Commission of the European Communities (DG XII), \\
Direction des Sciences de la Mati$\grave{\mbox{\rm e}}$re, CEA, France, \\
Bundesministerium f$\ddot{\mbox{\rm u}}$r Bildung, Wissenschaft, Forschung 
und Technologie, Germany,\\
General Secretariat for Research and Technology, Greece, \\
National Science Foundation (NWO) and Foundation for Research on Matter (FOM),
The Netherlands, \\
Norwegian Research Council,  \\
State Committee for Scientific Research, Poland, SPUB-M/CERN/PO3/DZ296/2000,
SPUB-M/CERN/PO3/DZ297/2000, 2P03B 104 19 and 2P03B 69 23(2002-2004)\\
JNICT--Junta Nacional de Investiga\c{c}\~{a}o Cient\'{\i}fica 
e Tecnol$\acute{\mbox{\rm o}}$gica, Portugal, \\
Vedecka grantova agentura MS SR, Slovakia, Nr. 95/5195/134, \\
Ministry of Science and Technology of the Republic of Slovenia, \\
CICYT, Spain, AEN99-0950 and AEN99-0761,  \\
The Swedish Natural Science Research Council,      \\
Particle Physics and Astronomy Research Council, UK, \\
Department of Energy, USA, DE-FG02-01ER41155. \\
EEC RTN contract HPRN-CT-00292-2002.\\



\end{document}